\documentclass[12pt,preprint]{emulateapj}
\usepackage{graphicx}

\shorttitle{Circumbinary Habitable Zones}
\shortauthors{Stephen R. Kane \& Natalie R. Hinkel}
\slugcomment{Submitted for publication in the Astrophysical Journal}

\begin{document}

\title{On the Habitable Zones of Circumbinary Planetary Systems}
\author{Stephen R. Kane, Natalie R. Hinkel}
\affil{NASA Exoplanet Science Institute, Caltech, MS 100-22, 770
  South Wilson Avenue, Pasadena, CA 91125, USA}
\email{skane@ipac.caltech.edu}


\begin{abstract}

The effect of the stellar flux on exoplanetary systems is becoming an
increasingly important property as more planets are discovered in the
Habitable Zone (HZ). The Kepler mission has recently uncovered
circumbinary planets with relatively complex HZs due to the combined
flux from the binary host stars. Here we derive HZ boundaries for
circumbinary systems and show their dependence on the stellar masses,
separation, and time while accounting for binary orbital
motion and the orbit of the planet. We include stability
regimes for planetary orbits in binary systems with respect to the
HZ. These methods are applied to several of the known circumbinary
planetary systems such as Kepler-16, 34, 35, and 47. We also
quantitatively show the circumstances under which single-star
approximations break down for HZ calculations.

\end{abstract}



\section{Introduction}
\label{intro}

The Habitable Zone (HZ) is broadly defined as the region around a star
where water can exist in a liquid state on the surface of a planet
with sufficient atmospheric pressure. The boundaries of the HZ are
usually calculated based upon the properties of the host star, with
assumptions regarding the response of the planetary atmosphere to
stellar flux. Attempts to quantify these boundaries have evolved
considerably since initial discussions on the topic by
\citet{hua59,hua60}. \citet{har79} provided estimates for the Solar
System and later models of runaway greenhouse for Venus were
incorporated into the calculations by \citet{pol71} and
\citet{kas88}. \citet{kas93} calculated detailed one-dimensional
(altitude) climate models and considered conditions whereby the
equilibrium would sway to a runaway greenhouse effect or to a runaway
snowball effect, thereby providing robust estimates of the HZ
boundaries for main sequence stars.

Interestingly, much of the work on the HZ preceded the discovery of
exoplanets. The exoplanet parameter-space, specifically orbital period
and planetary mass, is slowly being expanded by different
techniques. For example, the radial velocity (RV) technique has
explored well beyond the HZ for main sequence stars, allowing
analysis of HZ properties of these systems \citep{kan12a}. Exoplanet
systems tend to be dominated by Jovian planets, many of which are in
highly eccentric orbits which causes the planet to pass through the HZ
\citep{kan12b}. There have also been several notable discoveries of
super-Earths in the HZ, such as Gl~581d \citep{udr07} and Gl 667Cc
\citep{ang12,del12}. The transit method is also contributing to the
census of HZ planets. The Kepler mission is designed to explore the HZ
\citep{kal11} and has already found several which meet this criteria,
such as Kepler-20b \citep{bor12}.

One of the more interesting discoveries to result from the Kepler
mission is that of circumbinary planets, or planets that orbit the
center-of-mass of a central binary system. The first of these
announced was Kepler-16b \citep{doy11}, soon to be followed by
Kepler-34b \& Kepler-35b \citep{wel12}, and the multi-planet system
Kepler-47b,c \citep{oro12}. There have been several discussions
regarding the HZ for these systems, though these to-date approximate
the calculations as a single-star for the purposes of determining the
HZ boundaries \citep{qua12,oro12}.

Here we perform a more thorough analysis of the HZ boundaries for
circumbinary planetary systems. We consider the contributions of both
stars and the resulting position dependence of the HZ boundaries. We
thus determine the mass/separation ratio dependence of these
boundaries and include the orbital motion of the binary to produce a
time dependent map of the HZ. We apply these calculations to the known
Kepler circumbinary planetary systems of Kepler-16, 34, 35, and 47.


\section{Habitable Zones of Single Star Systems}

The HZ for single star systems defined by stellar irradiation has been
discussed by a number of sources, most notably by \citet{kas93}. The
boundaries of the HZ for main sequence stars have been further
quantified as a function of effective temperature
\citep{und03,sel07,jon10}. Note that there are additional influences
on the habitability such as the effects of tides \citep{bar09,hel11}
and orbital stability \citep{kop10}. Here we consider only the
irradiation effects on the HZ and briefly describe the essential
components which we expand upon in future sections.

One of the fundamental stellar properties which determines the extent
of the HZ is the luminosity of the host star, which is approximated as
\begin{equation}
  L_\star = 4 \pi R_\star^2 \sigma T_\mathrm{eff}^4
\end{equation}
where $\sigma$ is the Stefan-Boltzmann constant. The model adopted by
\citet{kas93} utilizes the principle of a planetary energy balance:
net incoming solar radiation must equal the net outgoing infra-red
(IR) radiation. This method determines the stellar flux, $S$, for
which the outgoing IR radiation is sustainable for various atmospheric
assumptions. The underlying implication here is that the stellar flux
received at the HZ boundaries is dependent on the amount of IR
radiation incident upon the upper atmosphere. Thus there is a
temperature dependence on the host star. Using the boundary conditions
of runaway greenhouse and maximum greenhouse effects at the inner and
outer edges of the HZ respectively \citep{und03}, the stellar flux at
these boundaries are given by
\begin{equation}
  S_\mathrm{inner} = 4.190 \times 10^{-8} T_\mathrm{eff}^2 - 2.139
  \times 10^{-4} T_\mathrm{eff} + 1.268
  \label{hzeqn1}
\end{equation}
\begin{equation}
  S_\mathrm{outer} = 6.190 \times 10^{-9} T_\mathrm{eff}^2 - 1.319
  \times 10^{-5} T_\mathrm{eff} + 0.2341 \ .
  \label{hzeqn2}
\end{equation}
The inner and outer edges of the HZ are then derived from the
following
\begin{eqnarray}
  r_\mathrm{inner} = \sqrt{ L_\star / S_\mathrm{inner} } \\
  r_\mathrm{outer} = \sqrt{ L_\star / S_\mathrm{outer} }
\end{eqnarray}
where the radii are in units of AU and the stellar luminosities are in
solar units. The dependence of these boundaries upon the wavelength of
the incident radiation is of critical importance when considering two
separate radiation sources.


\section{Habitable Zones of Multi-Stellar Systems}

In this section we generalize the above principles to multi-stellar
systems. The HZ of planets in S-type orbits (planets orbiting one
component of a stellar binary) has been considered in detail by
\citet{egg12} and applied to $\alpha$ Centauri B by \citet{for12}, now
known to host a terrestrial exoplanet \citep{dum12}. We address the
specific case of HZ boundaries for planets in P-type orbits, where the
planet orbits the center of mass of both stars.


\subsection{Combined Black-body Spectrum and Stellar Flux}
\label{blackbody}

The radiation received at any point in a binary system is a
combination of two blackbody spectral radiances. First we consider
Planck's law as a function of wavelength, $\lambda$:
\begin{equation}
  I(\lambda,T) = \frac{2 h c^2}{\lambda^5} \frac{1}{e^{hc/\lambda k T}
    - 1}
\end{equation}
where $T$ is the effective temperature. Note that this equation is per
steradian. The stellar flux received at a particular location may then
be expressed by the following equations for star 1 and star 2:
\begin{equation}
  S_1 = \frac{2 \pi h c^2}{\lambda^5} \frac{1}{e^{hc/\lambda k
      T_{\mathrm{eff},1}} - 1} \left( \frac{R_{\star,1}}{r_1}
  \right)^2
  \label{seqn1}
\end{equation}
\begin{equation}
  S_2 = \frac{2 \pi h c^2}{\lambda^5} \frac{1}{e^{hc/\lambda k
      T_{\mathrm{eff},2}} - 1} \left( \frac{R_{\star,2}}{r_2}
  \right)^2
  \label{seqn2}
\end{equation}
where $R_{\star,1}$ and $R_{\star,2}$ are stellar radii and $r_1$ and
$r_2$ are the distances from a given location to star 1 and star 2
respectively. The total flux received at this location is then $S =
S_1 + S_2$. Wien's Displacement Law ($\lambda_\mathrm{max}
T_\mathrm{eff} = 2.9 \times 10^{-3}$) may then be used to estimate the
equivalent effective temperature of a single energy source that would
produce the same energy flux. For the purposes of integrating the
total flux over all wavelengths, Equations \ref{seqn1} and \ref{seqn2}
reduce to the Stefan-Boltzmann law. However, it is important to
consider the wavelength dependence of the incident flux, as we will
see in later sections.

\begin{figure}
  \includegraphics[angle=270,width=8.2cm]{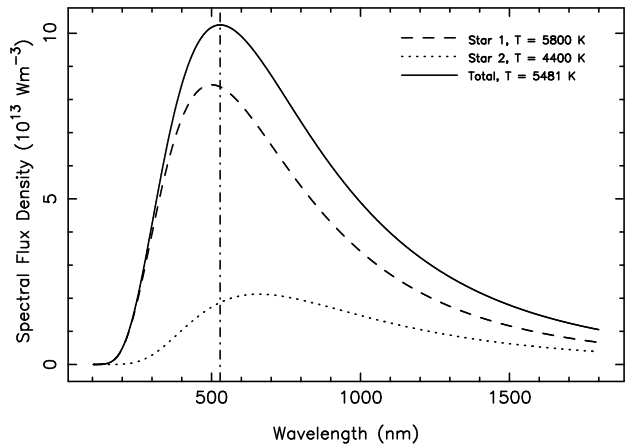}
  \caption{The blackbody spectra of the individual stars in the binary
    system and the combined flux for both stars. The peak of the
    combined spectrum peaks at a wavelength of 528~nm (vertical line)
    which, using Wien's Displacement Law, is equivalent to a
    temperature of 5481~K.}
  \label{bb}
\end{figure}

As an example, consider the case of a binary system consisting of a
G2V star and a K5V star, separated by 0.1~AU. Shown in Figure \ref{bb}
are the individual and combined blackbody spectra for this system. The
total flux peaks at a wavelength of 528~nm which is equivalent to a
temperature of 5481~K. This particular evaluation of the blackbody
spectra does not take into account the distances to the stars. When
the radii and distances are included, this introduces a spatial
dependence.

\begin{figure*}
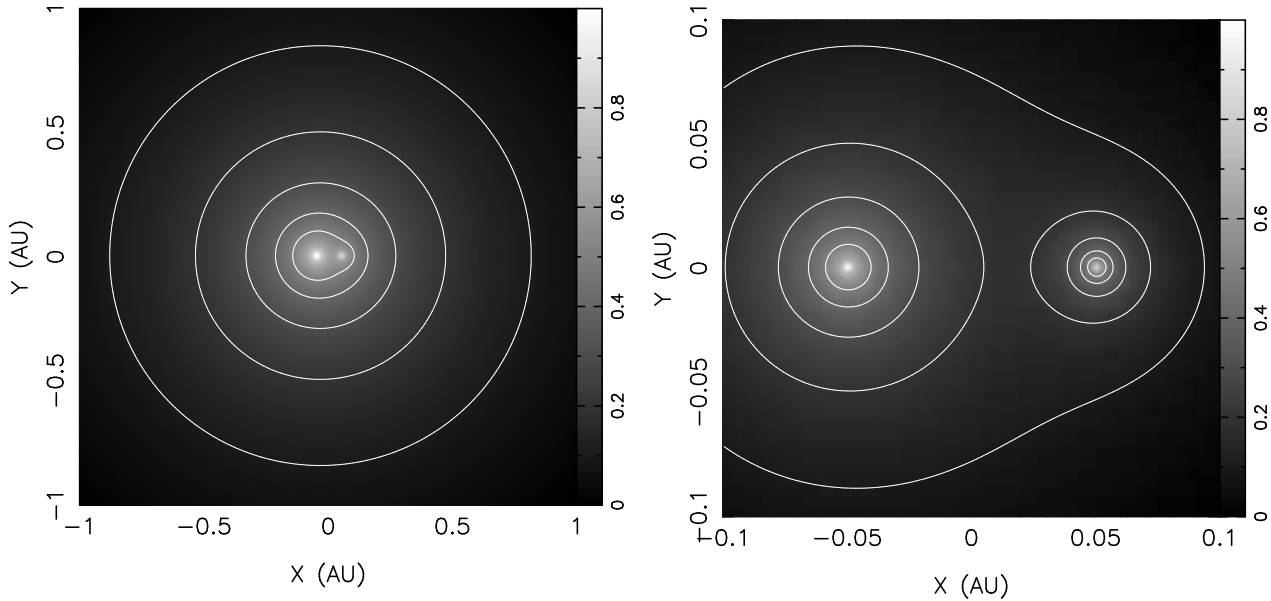

  \begin{center}
    \begin{tabular}{cc}
      \includegraphics[angle=270,width=8.2cm]{f02a.ps} &
      \includegraphics[angle=270,width=8.2cm]{f02b.ps}
    \end{tabular}
  \end{center}
  \caption{The stellar flux map, depicted with both gray scale and
    contours, of a binary system consisting of a G2V and a K5V star
    with a separation of 0.1~AU. The right panel is a zoom of the left
    panel. In each panel, the calculated flux has been converted to a
    logarithmic scale and normalized to lie between 0 and 1. On this
    scale, the contours are equal to values of 0.1 to 0.5 in steps of
    0.1 moving from the outer edge inwards.}
  \label{sf}
\end{figure*}

Figure \ref{sf} shows a flux map of the same system with two different
zoom levels to emphasize the contribution from both stars. For the
purposes of contrast, the calculated flux has been converted to a
logarithmic scale and normalized to lie between 0 and 1. On this flux
scale, the contours are equal to values of 0.1 to 0.5 in steps of 0.1
moving from the outer edge inwards. There is a clear asymmetry
associated with the presence of the second star which is particularly
noticeable at distances closer than 0.3~AU.


\subsection{Habitable Zone Boundaries}
\label{hzbound}

The traditional HZ due to stellar irradiance for a single star is
calculated as a function of both the stellar flux received and the
peak wavelength of the energy distribution. For a single star, this
has a radial dependence in a spherical geometry which allows an
analytical solution for the inner and outer HZ boundaries. For a
multi-stellar system, the locations of these boundaries is far more
complex, since both the stellar flux and peak wavelength have
non-radial dependencies due to the combinational effect produced from
the two stars at a particular location.

\begin{figure*}
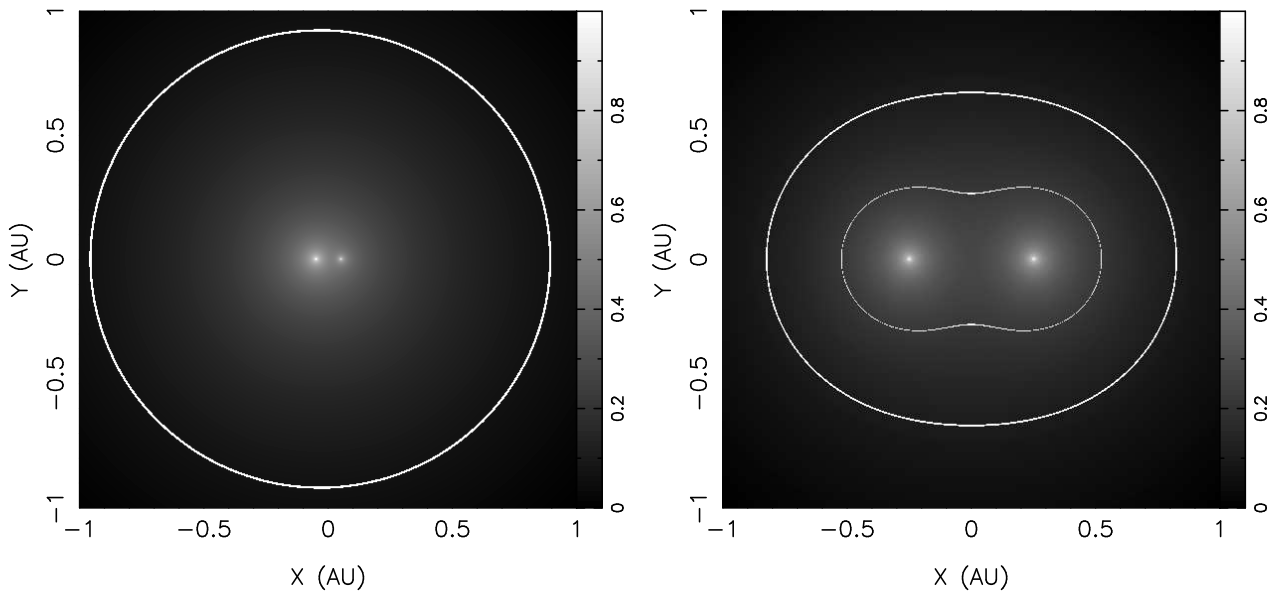

  \begin{center}
    \begin{tabular}{cc}
      \includegraphics[angle=270,width=8.2cm]{f03a.ps} &
      \includegraphics[angle=270,width=8.2cm]{f03b.ps}
    \end{tabular}
  \end{center}
  \caption{The stellar flux map shown as a gray scale map (as for
    Figure \ref{sf}) and the HZ boundaries for two different binary
    configurations. The left panel depicts these quantities for the
    G2V--K5V binary discussed in detail in Section \ref{blackbody}
    where only the inner edge of the HZ is shown on this scale. The
    right panel depicts these for an equal-mass binary consisting of
    two M0V stars separated by 0.5~AU. In this case both the inner and
    outer edge of the HZ are shown on the plot with a very clear
    asymmetry to those boundaries.}
  \label{hz}
\end{figure*}

We approach this problem by constructing a two-dimensional grid in the
orbital plane of the binary star. For each grid location we perform
the following steps: (1) calculate the combined stellar flux, (2)
calculate the combined blackbody function and determine the peak, (3)
convert the wavelength for the peak into effective temperature using
Wien's Displacement Law, and (4) calculate the expected stellar flux
at the HZ boundaries for this temperature using Equations \ref{hzeqn1}
and \ref{hzeqn2}. If the calculated stellar flux at that location
matches that expected for a HZ boundary then that grid position is
marked as such.

Figure \ref{hz} depicts the HZ boundaries for two example binaries:
the G2V--K5V binary described in \S \ref{blackbody} with a separation
of 0.1~AU (left panel), and an equal-mass binary consisting of two M0V
stars separated by 0.5~AU (right panel). The inner HZ boundary for the
G2V--K5V binary appears symmetric on this scale, but the distance of
the boundary from the primary (G2V) star varies between 0.90 and
0.94~AU. For the secondary star, this distance ranges from 0.84 to
1.01~AU. The asymmetric HZ boundaries for the M0V--M0V binary shown in
the right panel represents a far more extreme case where both the
inner and outer boundaries are plotted.

\begin{figure*}
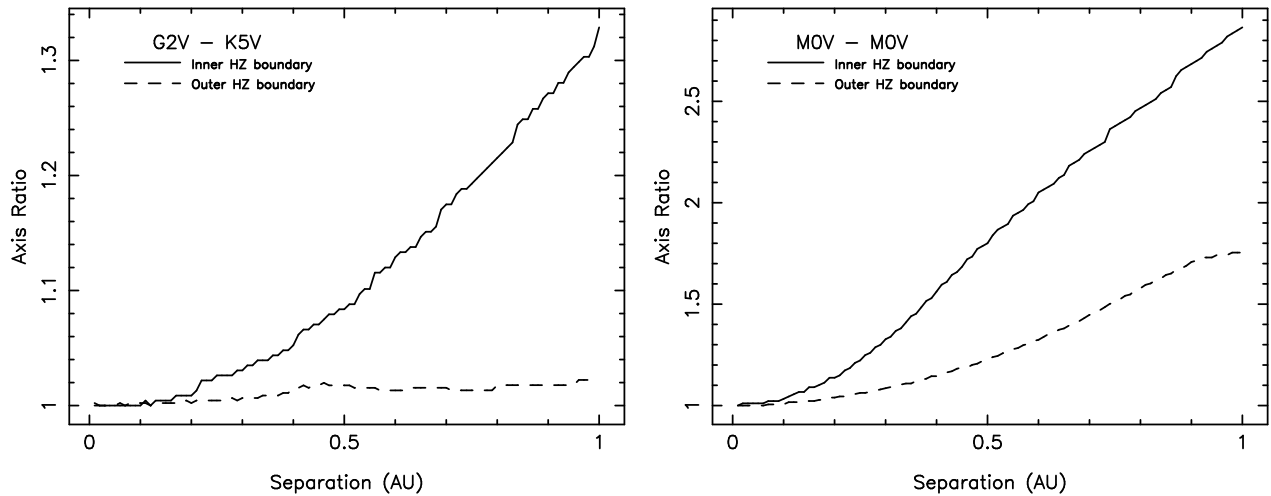

  \begin{center}
    \begin{tabular}{cc}
      \includegraphics[angle=270,width=8.2cm]{f04a.ps} &
      \includegraphics[angle=270,width=8.2cm]{f04b.ps}
    \end{tabular}
  \end{center}
  \caption{The dependence of the HZ asymmetry on the binary separation
    for the previously discussed cases of a G2V--K5V and M0V--M0V
    binary. At each separation, the axis ratio of both the inner and
    outer HZ boundaries are computed.}
  \label{rel}
\end{figure*}

The asymmetry in the HZ boundaries may be quantified as a function of
the binary separation by considering the axis ratio of both the inner
and outer boundaries of the HZ. We performed a simulation of this
dependence by calculating the HZ boundaries in the cases of the two
examples described above. We utilized a range of separations between
0.01 and 1.0~AU and located the minimum and maximum values of both the
inner and outer boundaries. The results of this simulation are shown
in Figure \ref{rel}. The case of the G2V--K5V shows moderate variation
in the inner HZ asymmetry but negligible variation in the outer
HZ. For the equal-mass M0V--M0V binary, the asymmetry rapidly
increases with increasing separation for both the inner and outer
boundaries.


\subsection{The Keplerian Orbit of the Binary}
\label{binaryorbit}

The orbital elements of binary stars are known to encompass a broad
range of orbital periods and mass ratios (see for example
\citet{hal05} and \citet{mat04}). This diversity results in an equally
varying range of HZ structures due to the mass ratios and binary
separations. From a fixed location relative to the binary
center-of-mass, the apparent HZ boundaries will have a time dependence
owing to the orbital motion of the binary.

For a Keplerian (non-circular) binary orbit, assuming a fixed
reference point, the changing HZ boundaries will have an additional
disparity resulting from the varying binary separation. This will
result in an oscillating HZ structure as the binary rotates around
their center-of-mass. As shown in the previous sections, the
oscillation will have a much larger effect for low-mass stars than for
solar-type stars. The large variation in flux within typical HZ
regions may ultimately render some binary systems uninhabitable.


\section{Planets Within the Habitable Zone}

Now we consider a planetary orbit which is passing through asymmetric
HZ regions and the flux/temperature variations that result.


\subsection{Stable Configurations}
\label{stable}

\begin{figure}
  \includegraphics[width=9.2cm]{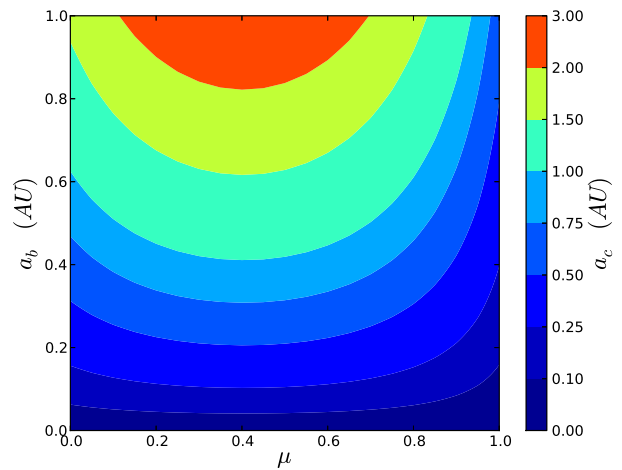}
  \caption{A contour plot of the dependence of the critical semi-major
    axis of a planetary orbit, $a_c$, on the mass-ratio, $\mu$, and
    separation, $a_b$ of a binary star \citep{hol99}.}
  \label{stability}
\end{figure}

\begin{figure*}
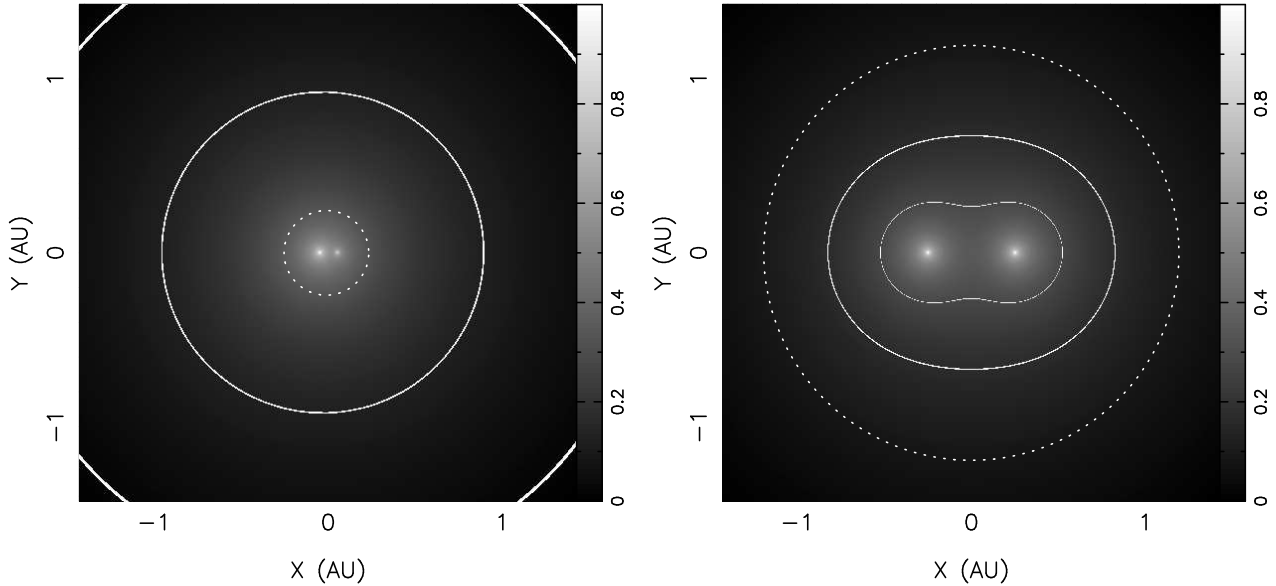

  \begin{center}
    \begin{tabular}{cc}
      \includegraphics[angle=270,width=8.2cm]{f06a.ps} &
      \includegraphics[angle=270,width=8.2cm]{f06b.ps}
    \end{tabular}
  \end{center}
  \caption{The stellar flux map shown as a gray scale map and HZ
    boundaries (solid lines) for a G2V--K5V (left panel) and M0V--M0V
    (right panel) binary, as for Figure \ref{hz}. Also shown in each
    case is the boundary for the critical semi-major axis (dotted
    line), beyond which planetary orbits may retain long-term
    stability. This shows that the stable planetary orbits may exist
    in the HZ for the G2V--K5V binary, but stable orbits are excluded
    for the HZ of the M0V--M0V binary.}
  \label{stab}
\end{figure*}

\begin{figure}
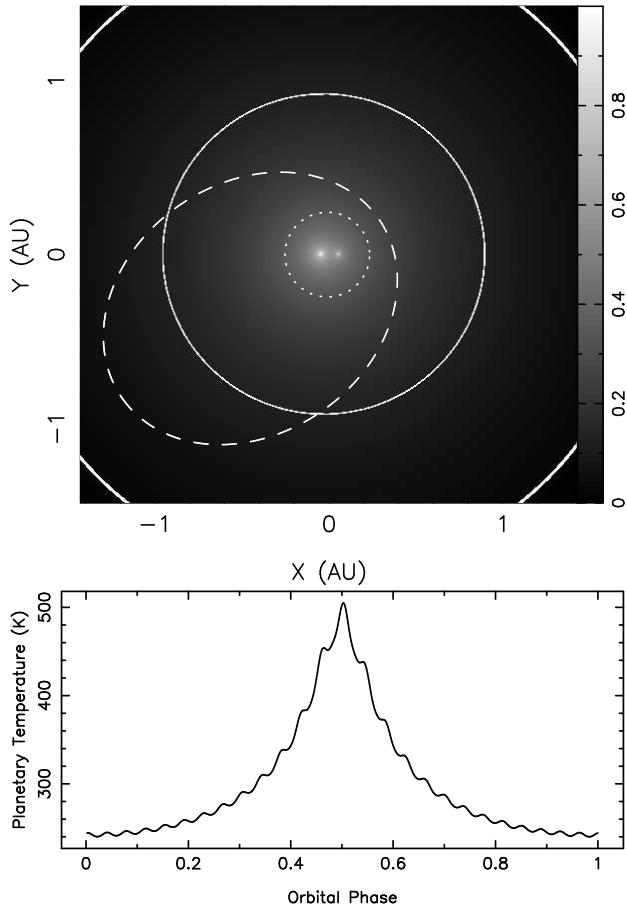

  \includegraphics[angle=270,width=8.2cm]{f07a.ps} \\
  \includegraphics[angle=270,width=8.2cm]{f07b.ps}
  \caption{Top panel: The G2V--K5V binary system showing the HZ
    boundaries (solid lines), critical semi-major boundary (dotted
    line), and a planetary orbit (dashed line). The planetary orbit
    has a semi-major axis of $a = 0.9$~AU and an eccentricity of $e =
    0.6$. Bottom panel: The equilibrium temperature of the planet as a
    function of planetary orbital phase. Phase 0.0 is at apastron and
    phase 0.5 is at periastron.}
  \label{orbit}
\end{figure}

Stable planetary orbits in both S-type and P-type systems have been
discussed by a variety of authors, including \citet{har77}, but more
recently by such authors as \citet{egg95} and \citet{mus05}. We adopt
the formalism of \citet{hol99} which provides analytical solutions for
planetary orbital stability in binary star systems. For P-type orbits,
this is a function of the binary separation, $a_b$, the mass ratio,
$\mu$, and the eccentricity of the binary orbit. The mass ratio is
defined as $\mu = m_2 / (m_1 + m_2)$, such that an equal mass binary
has a mass ratio of $\mu = 0.5$. For a binary in a circular orbit, or
$e = 0$, Equation 3 of \cite{hol99} may be expressed as
\begin{equation}
  a_c = (1.60 + 4.12 \mu - 5.09 \mu^2) a_b
\end{equation}
where $a_c$ is the minimum allowed semi-major axis of the planet.
Figure \ref{stability} shows the dependence on $a_b$ and $\mu$ of the
critical semi-major axis of a planet orbiting the center-of-mass of
the binary. For binary separations of $a_b < 0.2$~AU, the critical
semi-major axis is only weakly dependent upon the mass ratio of the
binary. As pointed out by \citet{hol99}, there often exists instability
islands beyond $a_c$ which occur at mean-motion resonances. Thus the
critical semi-major axis calculations here represent a lower limit on
the total stability regions within a given system.

Applying this stability criteria to the two examples discussed thus
far will yield whether or not a stable planetary orbit can be
maintained within the HZ of those systems. Figure \ref{stab} is
similar to Figure \ref{hz} in that it shows the stellar flux map and
HZ boundaries for the G2V--K5V and M0V--M0V binaries, but slightly
zoomed out so the outer part of the HZ for the G2V--K5V binary is
visible. Also shown are dotted lines which indicate the critical
semi-major axis for planets within the system, centered upon the
binary center-of-mass in each case. This clearly shows that stable
planetary orbits within the HZ for the G2V--K5V binary are possible
but the the presence of planets are excluded from the HZ of the
M0V--M0V binary.


\subsection{The Keplerian Orbit of the Planet}
\label{planetorbit}

A planet in orbit around the center-of-mass of the binary will
experience variable conditions due to the rotation of the binary, as
mentioned in \S \ref{binaryorbit}. An additional component of the
total flux variation is due to the Keplerian motion of the planetary
orbit. The radial dependence of the flux means that the influence of
the binary motion received at the planet will be more pronounced when
the planet is passing through periastron passage.

As a demonstration, we include the presence of a planet in the
G2V--K5V binary system example used throughout this paper (recall that
our particular M0V--M0V binary example excludes stable orbits within
the HZ for the binary separation used). We show the stellar flux map
(see \S \ref{blackbody}) in Figure \ref{orbit}, along with the HZ
boundaries (see \S \ref{hzbound}), the critical semi-major axis
boundary (see \S \ref{stable}), and the orbit of the planet. The
planet has a semi-major axis of 0.9~AU and an eccentricity of 0.6 such
that it spends most of the orbit within the HZ of the system. At each
location of the orbit, we recalculate the flux received by the planet
and estimate the equilibrium temperature of the planet assuming that
the incident flux is redistributed around the entire planet (see
\citet{kan11}). We also account for the stellar binary motion at each
location during the planetary orbit. The bottom panel of Figure
\ref{orbit} shows the change in this temperature during one complete
orbital phase as the planet moves from apastron (phase~$= 0.0$) to
periastron (phase~$= 0.5$) and back again. The long-term variation is
caused by the Keplerian orbit of the planet and the short-term
variations are caused by the 9 day orbital period of the binary. The
temperature variations can be dominated by either the orbits of the
binary or the planet, depending on the relative orbital
configurations.


\subsection{Binary/Planet Dynamical Interactions}

Thus far we have assumed a decoupling of the Keplarian motion of the
binary and the Keplerian motion of the planet. However, the secular
dynamical interactions in the system will result in a break-down of
the periodic behaviour one may otherwise obtain when you assume the
planet and binary are orbitally decoupled. Although the semi-major
axis of the planet's orbit will remain relatively unchanged, the shape
(eccentricity) and orientation (argument of periastron) will have
time-dependent components. This has been described in detail by
various authors, including analytical solutions for stellar triples
\citep{sod84,geo09}, single-star multi-planet systems \citep{lee03},
effects of general relativity \citep{egg06}, and the dynamics and
stability of planets in circumbinary orbits \citep{doo11}. The
amplitude of these variations depends on numerous factors, such as the
relative inclinations of the orbits \citep{far10}. Here we assume
co-planar orbits, as is the case for the Kepler systems discussed in
the following section.

The timescale of the secular dynamical interactions is largely
dominated by the periastron precession of the planet which occurs on
much shorter timescales than the precession of the binary. The
simulations by \citet{doo11} of test particles in circumbinary orbits
show that the period of the periastron precession of an orbiting
planet is related to the semi-major axis of the orbit, $a$, via the
power-law $\propto a^{3.5}$. Therefore, one can expect to be able to
measure the effects of periastron precession for systems similar to
those described here over the course of years, with a complete
periastron precession period in the range of several decades.
Although we are primarily concerned with the HZ of the binary systems,
the stability and secular dynamics of the planetary orbits in those
systems is an effect that one should consider when predicting the
percentage of the orbital phase within the HZ over long periods or
time.


\section{Application to Known Systems}
\label{application}

Here we apply the HZ boundary methods to several of the circumbinary
exoplanetary systems found via the Kepler mission.


\subsection{Kepler-16}

Kepler-16b was discovered by \citet{doy11} and was the first
circumbinary planet announced as a result of transit observations by
the Kepler mission. The binary system consists of 0.7 and 0.2
$M_\odot$ components in a 41 day eccentric orbit. The binary pair are
orbited by a 0.3~$M_J$ planet with an orbital period of 229 days. The
system has been the subject of numerous follow-up studies, such as
that carried out by \citet{win11} which showed that the plane of the
stellar orbit, the planetary orbit, and the primary’s rotation are all
closely aligned. Radial velocity observations by \citet{ben12} allowed
the dynamical determination of masses for both of the binary
components.

\begin{figure}
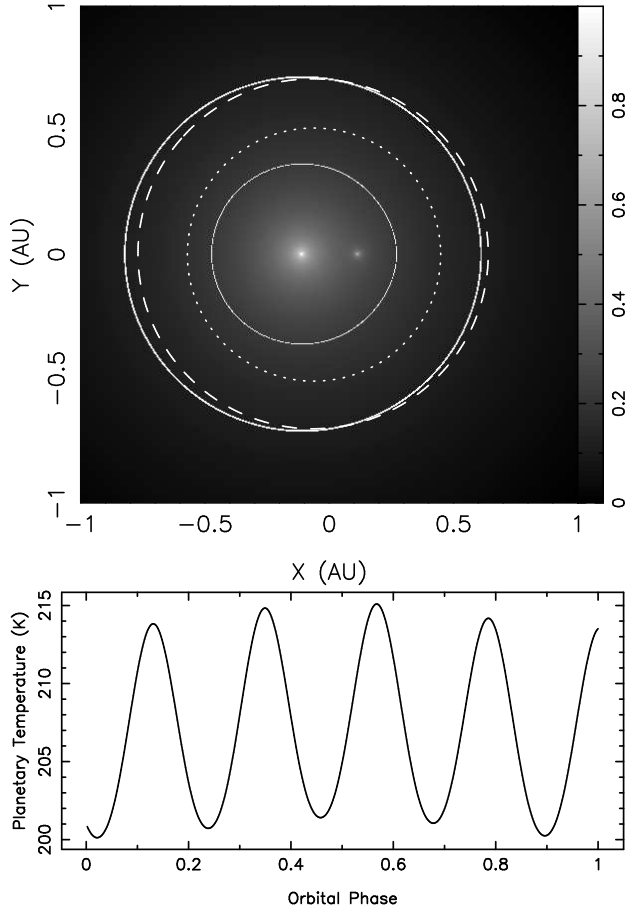

  \includegraphics[angle=270,width=8.2cm]{f08a.ps} \\
  \includegraphics[angle=270,width=8.2cm]{f08b.ps}
  \caption{Top panel: The Kepler-16 system showing the HZ boundaries
    (solid lines), critical semi-major boundary (dotted line), and
    planetary orbit (dashed line). Bottom panel: The equilibrium
    temperature of the planet as a function of planetary orbital
    phase. Phase 0.0 is at apastron and phase 0.5 is at periastron.}
  \label{kep16fig}
\end{figure}

Figure \ref{kep16fig} shows the flux map, HZ boundaries, critical
semi-major axis (stability) boundary, and planetary orbit for
Kepler-16. The effective temperature of the primary star,
$T_{\mathrm{eff},1} = 4450$~K, is provided by \citet{doy11}. For the
secondary star, we adopt the temperature of the spectral template star
proxy used by \citet{ben12}, GJ~905, of $T_{\mathrm{eff},2} =
2800$~K. Although there is a slight asymmetry in the inner HZ
boundary, this is inside the stability boundary. Overall, the flux is
clearly dominated by the primary star which produces an outer HZ
boundary centered on the primary. Our simulation shows that the offset
between the center-of-mass of the system and the flux distribution
results in the orbit of the planet moving in and out of the HZ during
one complete orbit. In other words, the asymmetry in the temperature
variations experienced by the planet can occur even for circular
orbits (see lower panel of Figure \ref{kep16fig}).


\subsection{Kepler-34 \& Kepler-35}

\begin{figure}
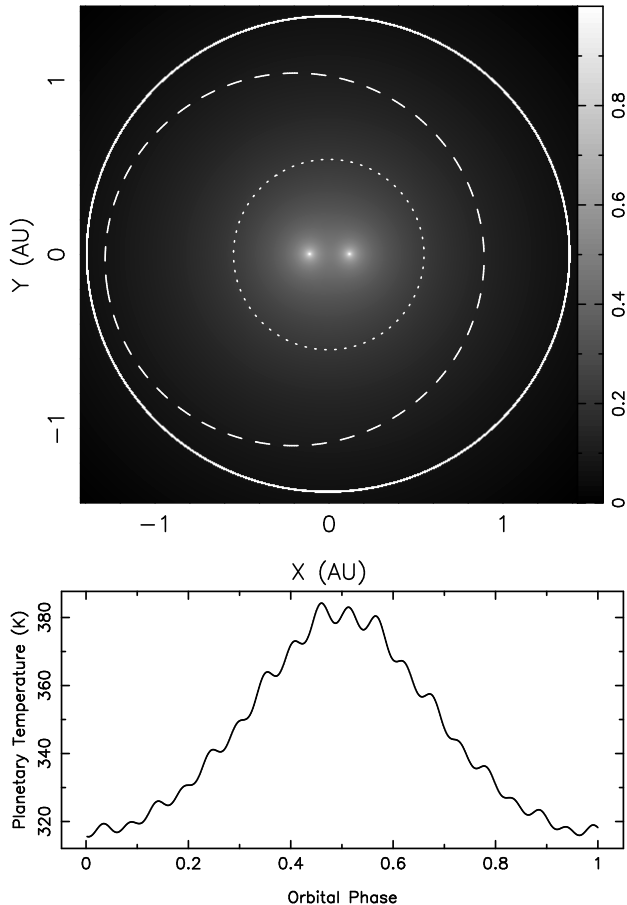

  \includegraphics[angle=270,width=8.2cm]{f09a.ps} \\
  \includegraphics[angle=270,width=8.2cm]{f09b.ps}
  \caption{Top panel: The Kepler-34 system showing the HZ boundaries
    (solid lines), critical semi-major boundary (dotted line), and
    planetary orbit (dashed line). Bottom panel: The equilibrium
    temperature of the planet as a function of planetary orbital
    phase. Phase 0.0 is at apastron and phase 0.5 is at periastron.}
  \label{kep34fig}
\end{figure}

\begin{figure}
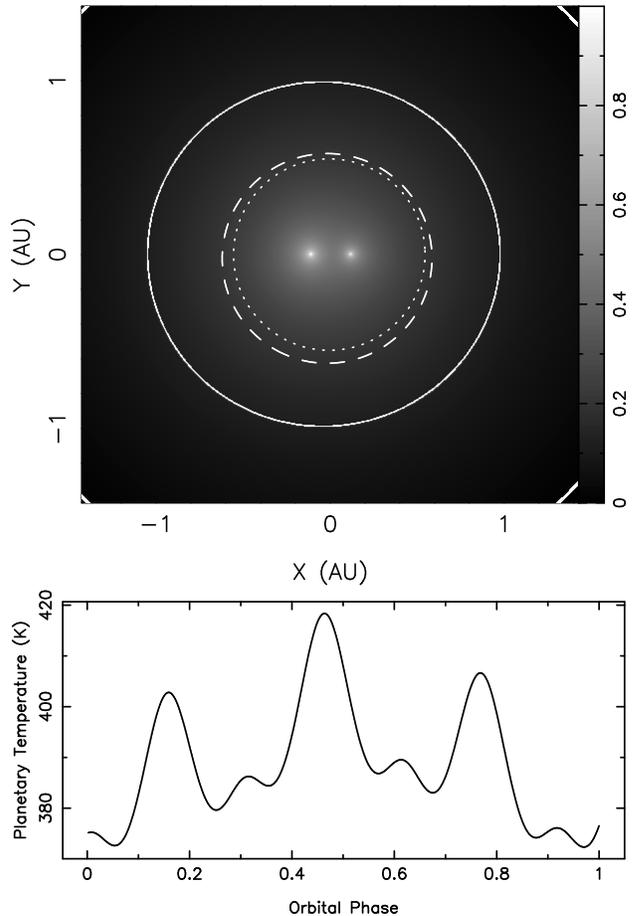

  \includegraphics[angle=270,width=8.2cm]{f10a.ps} \\
  \includegraphics[angle=270,width=8.2cm]{f10b.ps}
  \caption{Top panel: The Kepler-35 system showing the HZ boundaries
    (solid lines), critical semi-major boundary (dotted line), and
    planetary orbit (dashed line). Bottom panel: The equilibrium
    temperature of the planet as a function of planetary orbital
    phase. Phase 0.0 is at apastron and phase 0.5 is at periastron.}
  \label{kep35fig}
\end{figure}

\citet{wel12} announced the discovery of two circumbinary planetary
systems from the Kepler mission: Kepler-34 and Kepler-35. These
systems are quite comparable in terms of the mass ratio and separation
of the binary stars. Figures \ref{kep34fig} and \ref{kep35fig} show
our calculations for the HZ conditions in both systems, respectively.
The similarity of the stellar components in both cases yield
analogous centroids for the HZ and stability boundaries, unlike the
case of Kepler-16. The Kepler-34 binary consists of two near-solar
components with $T_{\mathrm{eff},1} = 5913$~K and $T_{\mathrm{eff},2}
= 5867$~K separated by 0.23~AU. Only the inner HZ boundary is shown
since the combined flux of these stars pushes the outer boundary much
farther out than for our own Solar System. The planet is in an
eccentric orbit inside of the inner HZ boundary with a semi-major axis
of 1.09~AU. The resulting planetary temperature variation has both
clear short-term and long-term components.

By comparison, the Kepler-35 system consists of stellar components
with effective temperatures of 5606~K and 5202~K, respectively,
separated by 0.18~AU. The HZ boundaries are thus smaller in size
compared with Kepler-34. The planet's near-circular orbit is very
close to the stability boundary of the system, therefore the
binary. This causes the temperature profile of the planet to be very
sensitive to the orbital motion of the binary, as can be seen in the
plot of the temperature variations during one complete orbital period.


\subsection{Kepler-47}

\begin{figure}
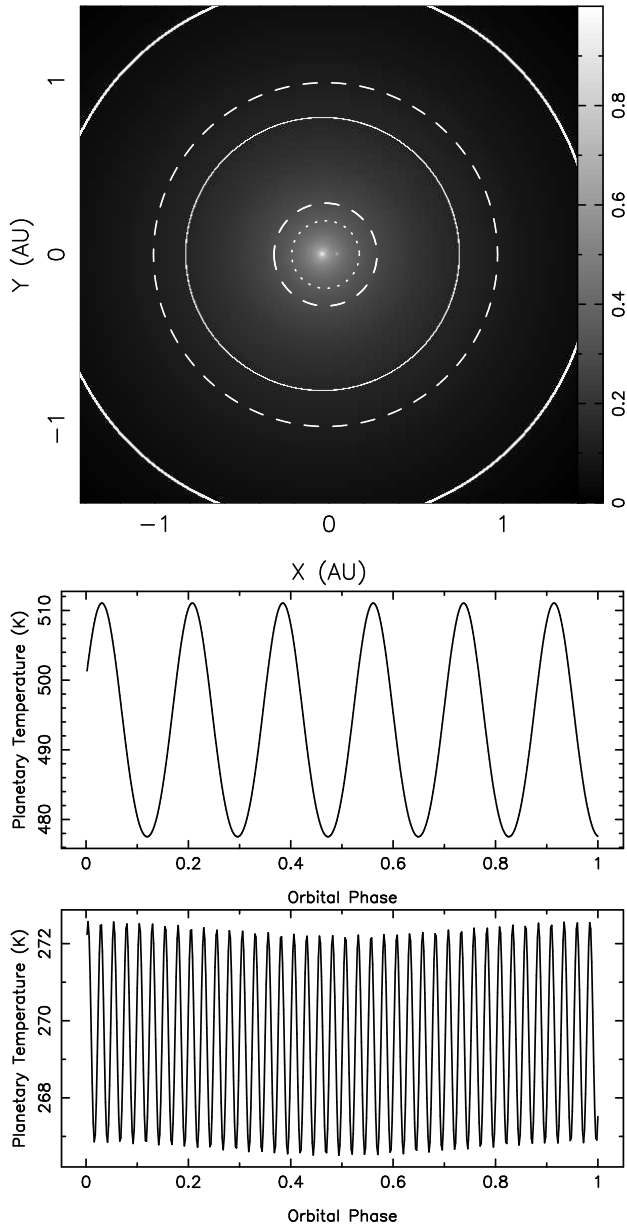

  \includegraphics[angle=270,width=8.2cm]{f11a.ps} \\
  \includegraphics[angle=270,width=8.2cm]{f11b.ps} \\
  \includegraphics[angle=270,width=8.2cm]{f11c.ps}
  \caption{Top panel: The Kepler-47 system showing the HZ boundaries
    (solid lines), critical semi-major boundary (dotted line), and
    orbits for the b and c planets (dashed lines). Middle panel: The
    equilibrium temperature of the b planet as a function of planetary
    orbital phase.  Bottom panel: The equilibrium temperature of the c
    planet as a function of planetary orbital phase. Phase 0.0 is at
    apastron and phase 0.5 is at periastron.}
  \label{kep47fig}
\end{figure}

The Kepler-47 system, discovered by \citet{oro12}, contains a binary
star whose components are quite different, similar to Kepler-16. The
two stars have temperatures of 5636~K and 3357~K, respectively, and
are separated by only 0.08~AU. The resulting stellar flux dominated by
the primary produces asymmetry in the flux only in the vicinity of the
binary (within the stability boundary). The main difference between
Kepler-47 the other studied Kepler systems, however, is that this
system has two known transiting planets. Figure \ref{kep47fig} shows
our calculations for the Kepler-47 system, as well as the orbits and
temperature profiles for the two planets. Both planets are in
near-circular orbits with semi-major axes as 0.30~AU and 0.99~AU,
respectively. As noted by \citet{oro12}, the outer planet resides
within the HZ of the system throughout the entire orbit. Since the
planets are in non-eccentric orbits, the temperature variations are
dominated by the binary orbit, with an additional variation caused by
the orbital center-of-mass offset from the primary stellar flux. The
amplitude of the variations for the outer planet are relatively small,
$\sim 5$~K, which is negligible for habitability considerations in this
region.


\section{Conclusions}

Calculating the amount of time an exoplanet spends within the HZ
around a circumbinary system is relatively complicated. First, we must
consider the amount of stellar flux from both stars received by the
exoplanet at any point in its orbit (\S \ref{blackbody}). Then we must
determine the wavelength at which this combined flux peaks and how
that translates into an effective temperature (\S
\ref{blackbody}). This will then allow us to determine the inner and
outer boundaries of the HZ (\S \ref{hzbound}). Depending on the
configuration of the binary masses and separation, the HZ boundaries
may then experience a significant time-dependent orbital motion with
respect to the rotation of the stars (\S \ref{binaryorbit}).  Finally,
the stability of the binary system must be analyzed (\S \ref{stable})
in order to fully understand the planetary orbit with respect to the
oscillating HZ (\S \ref{planetorbit}).

We have applied our analysis to a number of known exoplanet-hosting
binary systems (\S \ref{application}), namely Kepler-16, 34, 35, and
47. We have demonstrated that the circumstances under which a
single-star approximation is viable is solely reliant on the stellar
masses and separation of the binary. For example, the components in
Kepler-16 and Kepler-47 are so dominated by the primary star that our
analysis does not significantly change the approximation of the HZ as
compared to a single-star model. However, the HZ in Kepler-16 is
offset from the center of mass of the binary resulting in a planetary
orbit that alternates in and out of the boundary. The binary stars
within Kepler-34 and Kepler-35 are similar in mass and at such
distances from each other that require our analysis to determine
accurate HZs. The benefit of analyzing these Kepler systems, and
Kepler systems in general, is that they exemplify the diversity we
expect when exoplanets orbit binary systems.

To date, these circumbinary planetary systems are second only in
complexity to the recently discovered double-binary system with an
interposed exoplanet \citep{sch12}. In this instance, an exoplanet
revolves around an eclipsing binary in a P-type configuration with a
period of $\sim$ 138 days, while a second visual binary orbits at
$\sim$ 1000 AU. The method that we have described here for determining
the HZ boundaries could be applied to this quadruple system, as well
as any system with an $n$-number of stars. The flux from the $n$-stars
would need to be combined and the effective temperature
determined. However, whether an exoplanet would be able to maintain a
stable configuration within that system would be another matter. We
anticipate further discoveries by the Kepler mission of new and
interesting stellar and planetary configurations on which to apply our
method.


\section*{Acknowledgements}

The authors would like to thank the anonymous referee, whose comments
greatly improved the quality of the paper. The authors acknowledge
financial support from the National Science Foundation through grant
AST-1109662.


\end{document}